\documentstyle[aps,prl,twocolumn,epsf,amsmath,graphicx,epsfig]{revtex}

\newcommand\ltdash{\raise-1.8pt\hbox{$\scriptscriptstyle |$}}

\def\figx#1#2{\includegraphics[width=#1]{#2}}

\newcommand{\dg}{^{\dagger}}

\newcommand{\la}{\langle}
\newcommand{\ra}{\rangle}

\newcommand{\rarrow}{\rightarrow }

\newlength{\bxwidth}

\newlength{\fight}
\newcommand \beq  {\begin{equation}}
\newcommand \eeq  {\end{equation}}
\newcommand \bea {\begin{eqnarray} }
\newcommand \eea {\end{eqnarray}}

\newcommand{\figo}[3]{
\begin{figure}[here]
\[ 
\figx{\fight}{#1}
\] 
\vspace*{-2mm}
\caption{\label{#3}
\small
{#2}
}\end{figure} 
\noindent }
\begin{document}
\draft
\title{What is the fate of the heavy electron at a quantum critical
point?  }

\author{$^{1}$ P. Coleman and $^{2}$
C. P{\'e}pin}

\address{$^{1}$Center for Materials Theory,
Department of Physics and Astronomy, 
Rutgers University, Piscataway, NJ 08854, USA.}

\address{$^{2}$SPhT, 
L' Orme des Merisiers, 
CEA-Saclay, 
91191 Gif-sur-Yvette, France.}

\vskip 0.2truein
\address{\mbox{ }}
\address{\mbox{ }}
\address{\parbox{14cm}{\rm \mbox{ }\mbox{ }
A growing body of evidence suggests that the quantum critical behavior at
the onset of magnetism in heavy fermion systems can not be understood in
terms of a simple quantum spin density wave.  This talk will discuss the
consequences of this conclusion, touching on its possible implications
in the realm of two dimensional systems and outlining current theoretical
and experimental efforts to characterize the nature of the critical point
in heavy fermion materials.}}
\address{\mbox{ }}
\address{\parbox{14cm}{\rm PACS No: }}
\maketitle

\makeatletter
\global\@specialpagefalse
\makeatother

\narrowtext

\par
\vskip 0.2truein

\subsection{Introduction.}\label{} Discoveries over the past decade
have brought a new awareness of
the importance of quantum critical points in condensed matter
physics. A quantum critical point (QCP) is a zero-temperature 
instability between two phases of matter
where quantum fluctuations develop long range
correlations in both space and time\cite{sachdevbook}.  These special points exert
wide-reaching influence on the finite temperature properties 
of a material. 
Systems close 
to quantum criticality develop a new excitation structure, 
they display novel thermodynamic, transport and magnetic
behavior. They
also 
have marked 
a predeliction towards the
development of new kinds of order, such as anisotropic
superconductivity.
A dramatic 
example is provided by the cuprate superconductors. By doping with holes, 
these materials pass through one or more quantum phase transitions: 
from a  Mott insulator to a 
\begin{figure}[here]
\[ 
\figx{\fight}{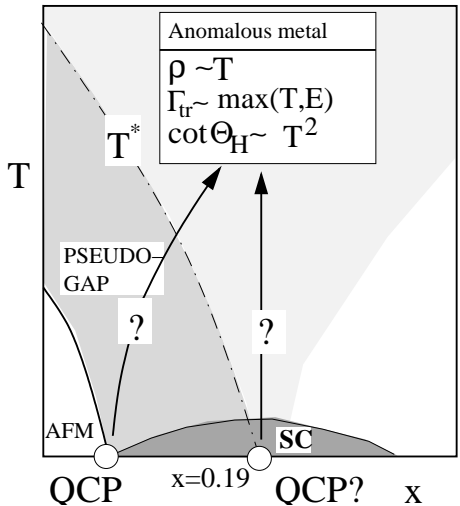}
\] 
\vspace*{-2mm}
\caption{\label{fig1}
\small
Schematic phase diagram for cuprate superconductors showing
location of possible quantum critical points.  
One of these QCP may be responsible for the anomalous normal state
which develops above the pseudogap scale. 
}\end{figure} 
\noindent 
metal with a spin gap at low doping, and
at
higher doping 
a second QPT 
may occur when the spin gap closes 
\cite{loram} (Fig. 1). 
The singular interactions induced by quantum criticality 
may be
the driving force for both the high temperature
superconductivity and 
the anomalous metallic state 
above the spin gap temperature $T^{*}$.\cite{castellani}

This paper discusses quantum criticality in heavy fermion
materials.  These materials
offer a unique opportunity to study quantum
criticality under controlled conditions.  By the application of
pressure or doping, heavy fermion materials can be tuned
through a quantum critical point from a metallic
antiferromagnet into a paramagnet (Fig. 2). 
Unlike the cuprate metals, here the 
the paramagnetic phase is a well characterized Fermi
liquid,\cite{landau,early,early2}   with heavy Landau
quasiparticles, or ``heavy electrons''.
A central property of these quasiparticles, is the existence of a
finite overlap ``$Z$'' between a
single quasiparticle state, denoted by $\vert \hbox{qp}^{-}\ra$ and
the state formed by adding a single electron to the ground-state,
denoted by $\vert e^{- }\ra = c\dg _{{\bf k }\sigma}\vert 0\ra$.
This quantity is closely related
to the ratio $m/m^{*}$ of the electron to quasiparticle mass,  
\begin{equation}\label{}
Z= \vert \la e^{- }\vert \hbox{qp}^{-}\ra \vert ^{2}\sim \frac{m}{m^{*}}.
\end{equation} 
A wide body of evidence suggests that $m^{*}/m$
diverges at a heavy fermion QCP, indicating
that 
\[
Z\rarrow 0 \qquad \qquad  (\hbox{$P\rarrow P_{c}$}).
\]
The state which forms at the QCP 
is referred to as a ``non-Fermi'' or ``singular Fermi
liquid''. \cite{varma2001,questions}
By what mechanism does this 
break-down in the Landau quasiparticle occur? 
\subsection{Properties of the Heavy Fermion Quantum Critical Point}

There is a  growing list of heavy fermion systems that
have been tuned to an antiferromagnetic QCP
by the application of pressure or by doping (Table 1.). 
These materials display many common properties

\vspace{0.1 truein}
{\bf Table. 1. Selected Heavy Fermion compounds with quantum critical points.}
{\small
\vskip 0.1 truein
\begin{tabular}{|c||c|c|c|c|}
\hline
Compound & 
$x_c$/$P_c$/$H_c$ & $\frac{C_v}{T}\rarrow \infty $? & $\rho\sim T^a $ & Ref.\\
\hline
$YbRh_2Si_{2-x}Ge_x$ & $x_c=0.1$ &
$Log \left( \frac{T_o}{T}\right)$
$^{(b)}$ & $ T$ & \cite{steglich} \\
\hline
$CeCu_{6-x}Au_x $ & $x_c=0.1$ & $Log \left( \frac{T_o}{T}\right)$ & $ T + c $ & \cite{hilbert1,hilbert2,schroeder} \\
\hline
$CeCu_{6-x}Ag_x$ & $x_c=0.09$ & $Log \left( \frac{T_o}{T}\right)$ & $T^{1.1}$ & \cite{heuser}\\
\hline
\hline
$CeNi_2Ge_2$ & $P_c =0$  & $Log \left( \frac{T_o}{T}\right)$$^{(a)}$
& $T^{1.2}$ & \cite{grosche,gegenwart,julian,braithwaite,newgegen} \\
\hline
$U_2Pt_2In$ & $P_c=0$  & $Log \left( \frac{T_o}{T}\right)$
 & $T$ & \cite{devisser}\\
\hline
$U_2Pd_2 In$ & $P_c<0 $ &  ? & $T + c $ & \cite{devisser} \\
\hline
$CePd_2Si_2$ & $ P_c>0$  & ? & $ T^{1.2}$ & \cite{grosche} \\
\hline
$CeIn_3$ & $P_c>0$ & ? & $T^{1.5} $ & \cite{grosche,mathur} \\
\hline
$U_3Ni_3Sn_4$ & $P_c >0$ & no & ? & \cite{devisser2} \\
\hline
$CeCu_2Si_2$ & $P_c=0$ & no & $T^{1.5}$ & \cite{cecu2si2}\\
\hline
\hline
$CeRu_2Si_2$ & $H^{\parallel}_c=7.7T$ &  $Log \left( \frac{T_o}{T}\right)$
 &  & \cite{aoki2}\\
\hline
$UPt_3$ & $H^{\perp}_c=20T$ &  $Log \left( \frac{T_o}{T}\right)$
 & $T^{1.2}$ & \cite{stewartupt3}\\
\hline
$Sr_3Ru_2O_7$ & $H^{\parallel}_c=7.7T$ &  $Log \left( \frac{T_o}{T}\right)$
 & $T^{1.1}$ & \cite{mackenzie}\\
\hline
\hline
\end{tabular}
}

\vspace{0.1 truein}
{\small $^{(a)}$ New  data\cite{newgegen} show a stronger divergence at lower
temperatures, and $\gamma \sim A- B \sqrt{T}$ at intermediate temperatures.

\noindent $^{(b)}$ At low temperatures, $\gamma$ diverges more rapidly than 
 $Log \left( \frac{T}{T_o}\right)$\cite{steglich}.
}\\ 

\begin{itemize}

\item {\bf Fermi liquid behavior in the paramagnet}, as indicated by
the emergence of a quadratic temperature dependence in the resistivity
in the approach to the QPT $\rho =\rho_{o}+A T^{2}$
\cite{devisser2,flouquet} at ever lower temperatures.

\item 
{\bf  Divergent  specific heat} 
at the QCP, typically with 
a logarithmic temperature dependence, 
\begin{equation}\label{}
\gamma (T) = \frac{C_{v} (T)}{T} = \gamma _{0} \log \left[\frac{T_{o}}{T} \right],
\end{equation}
suggesting that the Fermi temperature vanishes
and the quasiparticle effective masses diverge 
\begin{equation}\label{}
T_{F}^{*}\rarrow 0, \qquad \frac{m^{*}}{m}\rarrow \infty 
\end{equation}
at the QCP.
Further 
support for this conclusion is provided by the observation that
the quadratic coefficient $A$ of  the resistivity grows, and may diverge
in the approach to the quantum critical point\cite{devisser}.

\item {\bf  Quasi-linear resistivity}
\begin{equation}\label{}
\rho \propto T^{1+\epsilon},
\end{equation}
at the QCP with
$\epsilon$ in the range $ 0-0.6$.
In critical $YbRh_{2}Si_{2-x}Ge_{x}$, $\rho \propto T$
over three decades\cite{steglich}.

\item {\bf  Non-Curie spin susceptibilities}
\begin{equation}\label{}
\chi ^{-1} (T)= \chi _{0}^{-1}
+ c T^{a} 
\end{equation}
with $a<1$ observed in critical 
$Ce Cu_{6-x}Au_{x}$ (x=0.1), $YbRh_{2} Si_{2-x}Ge_{x}$ (x=0.1)
and $CeNi_{2}Ge_{2}$.

\item {\bf $E/T$ and $H/T$ Scaling.} 
In critical $Ce Cu_{6-x}Au_{x}$ and $YbRh_{2} Si_{2-x}Ge_{x}$ the differential magnetic
susceptibility
$dM/dH$
exhibits $H/T$ scaling, 
\begin{equation}\label{}
(dM/dH)^{-1} = \chi _{0}^{-1}
+ c T^{a} g[H/T], 
\end{equation}
where $a\approx 0.75$.
Neutron measurements\cite{schroeder}
show $E/T$ scaling\cite{varma1989,aronson} in 
the dynamical spin susceptiblity of critical $Ce Cu_{6-x}Au_{x}$, 
throughout the Brillouin zone, parameterized in the form
\begin{equation}\label{lab1}
\chi^{-1} ({\bf q},\omega ) =  T^{a}f (E/T)+\chi _{0}^{-1} ({\bf q})
\end{equation}
$F[x]\propto (1-ix)^{a}$. 
Scaling behavior
with a single anomalous exponent in 
the momentum-independent
component of the dynamical spin susceptibility 
suggests an emergence of {\sl local} magnetic moments  which 
are {\sl critically correlated in time} at the quantum critical
point\cite{schroeder}. 
\end{itemize} 

\begin{figure}
\[ 
\figx{\fight}{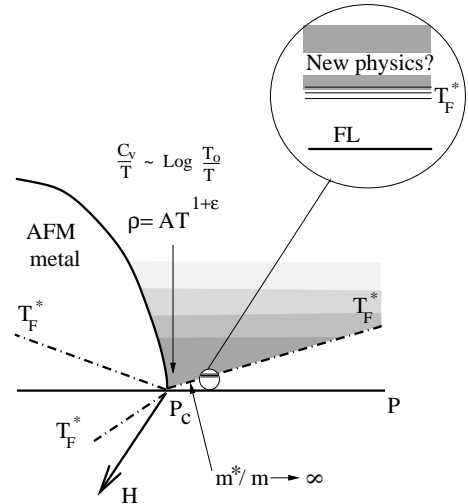}
\] 
\vspace*{-4mm}
\caption{\label{fig1}
\small
Illustration of quantum critical physics
in heavy fermion metals. 
As criticality is 
approached from either side of the transition, the temperature 
scale $T_{F}^{*}$ on which Fermi liquid behavior breaks down goes to zero.
A key challenge is to characterize the 
new class of universal excitations which develops above 
$T_{F}^{*}$.
}\end{figure} 
\noindent

Growing evidence suggests a strong 
connection between the heavy fermion QCP
and ``metamagnetism''. 
A number of strongly correlated electron systems, such as 
$CeRu_{2}Si_{2}$\cite{aoki2} and
$UPt_{3}$\cite{stewartupt3}, and 
$Sr_{3}Ru_{2}O_{7}$\cite{mackenzie}
exhibit a sudden rise 
in the magnetic polarization 
at a finite ``critical'' field. 
Properties characteristic of an
heavy fermion QCP,  (such as the logarithmic
dependence of the specific heat and a quasi-linear resistivity)
are seen to develop at such metamagnetic transitions. 
\noindent 
In all likelihood, this reflects the close
vicinity to a  {\sl ferromagnetic} QCP (Fig 3), but why 
is the physics similar to that of a heavy fermion QCP? 
\fight=0.8\columnwidth
\figo{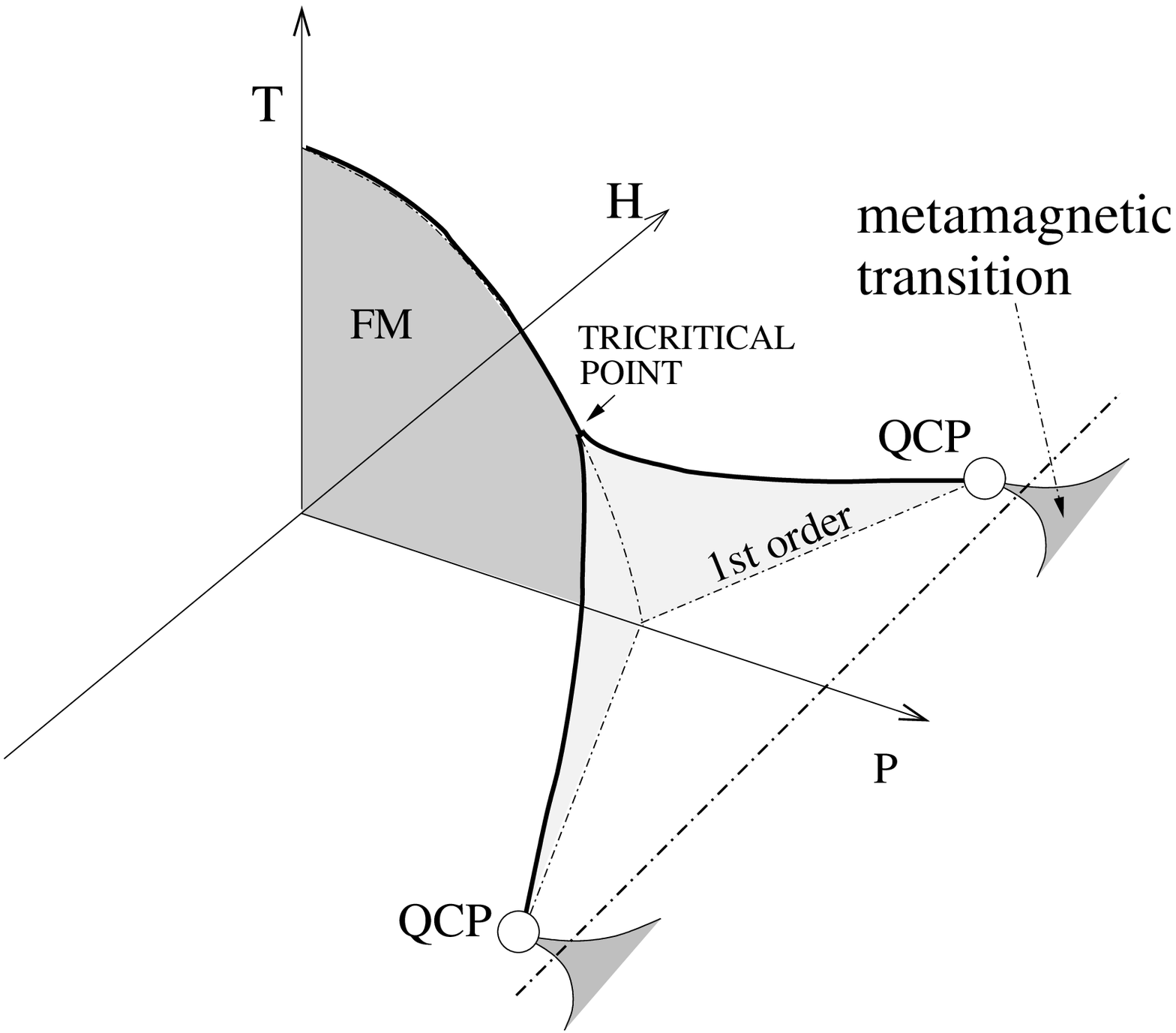}{Schematic phase diagram for a metamagnet. Second order
lines radiate from the tricritical point[29], intersecting the 
$T=0$ plane at quantum critical points.  Ambient pressure is denoted
by a dotted line.  When the material passes close to the QCP, 
a metamagnetic transition takes place. 
}{fig3}
\vskip -0.4truein
\subsection{Universality}

Usually, the physics of a metal above its
Fermi temperature depends on the 
detailed chemistry and
band-structure of the material: it is non-universal. 
However, if the renormalized Fermi temperature $T^{*}_{F}
(P)$ can be tuned to become arbitrarily small compared with the 
characteristic scales of the material as one approaches a QCP, 
we expect that 
the ``high energy'' physics
{\sl above} the Fermi temperature $T^{*}_{F}$ is itself, 
\underline{universal}. 

Quantum critical behavior implies a divergence of
the long distance and long-time correlations in the material. 
Finite temperatures introduce 
the cutoff timescale
\begin{equation}\label{}
\tau_{T} = \frac{\hbar }{k_{B}T}
\end{equation}
beyond which coherent quantum processes  are dephased by thermal
fluctuations. Renormalization group principles\cite{hertz}
imply that the quantum critical physics has an 
upper-critical dimension $d_{u}$. For $d<d_{u}$, the $\tau _{T}$
becomes \underline{the }
correlation time $\tau $ of the system\cite{zinnjustin}, so frequency dependent
correlation functions and response functions take the form 
\begin{equation}\label{}
F (\omega,T)= \frac{1}{\omega^{\alpha }} f (\omega \tau_{T} )= 
\frac{1}{\omega^{\alpha }} f (\hbar \omega /k_{B}T).
\end{equation}
leading to 
$E/T $
scaling\cite{sachdev}.
By contrast, for $d>d_{u}$ the correlation
time is sensitive to the details of the short-distance interactions
between the critical modes, and in general  $\tau ^{-1}\propto T^{1+b}$
Thus $E/T$ scaling with a non-trivial exponent strongly suggests that 
the underlying physics of the heavy fermion quantum critical point is 
governed by universal physics with $d_{u}> 3$.

%
%
%
%
%

\subsection{Failure of the Spin Density Wave picture }

The commonly accepted picture of the heavy fermion QCP
assumes the non-Fermi liquid behavior derives from Bragg diffraction of
the electrons off a quantum-critical spin density wave (QSDW)\cite{hertz,paladium,moriya,millis}. 
The virtual emission  of these soft fluctuations,
\begin{equation}\label{}
e^{-} \rightleftharpoons e^{-}+ \hbox{spin fluctuation}
\end{equation}
generates 
a retarded interaction 
\begin{equation}\label{}
V_{eff} ({\bf q},\omega )= g^{2}\overbrace {\left[ \frac{\chi _{0}}{({\bf q}-{\bf
Q})^{2}+\xi^{-2}-\frac{i\omega }{\Gamma_{\bf Q}}}\right]}^{{\chi ({\bf
q},\omega)} }
\end{equation}
between the electrons, 
where $\chi ({\bf q},\omega )$
is the dynamical 
spin susceptibility of the collective modes. 
The damping term $-i\omega/\Gamma_{{\bf  Q}}$
of the magnetic fluctuations is derived from
the linear density of particle-hole states in the Fermi sea.
$\xi^{-1}
$ and 
$\tau ^{-1}= \Gamma
_{{\bf Q}}\xi^{-2}$ are the the inverse spin correlation length and
correlation times respectively. 
In real space, 
\begin{equation}\label{}
V_{eff} (r,\omega =0)\propto \frac{e^{-r/\xi}}{r}e^{i {\bf Q}\cdot {\bf r}}
\end{equation}
is a  ``modulated ''
Yukawa potential whose range $\xi\sim
(P-P_{c})^{-\frac{1}{2}}
\rightarrow\infty  $ at the QCP. 
Unlike a ferromagnetic QCP, 
the modulated potential only  effects electron quasiparticles along 
``hot lines'' on the Fermi surface, 
that are separated by the wave-vector $\bf Q$ and satisfy
$\epsilon_{\bf k}= \epsilon_{{\bf  k}+{\bf  Q} }$. At a finite
temperature, electrons within a momentum range  $\sim \sqrt{T}$
are affected by this critical scattering (Fig. 4.). This limits
the ability of this singular potential to generate non Fermi liquid
behavior. 
\fight=0.8\columnwidth
\figo{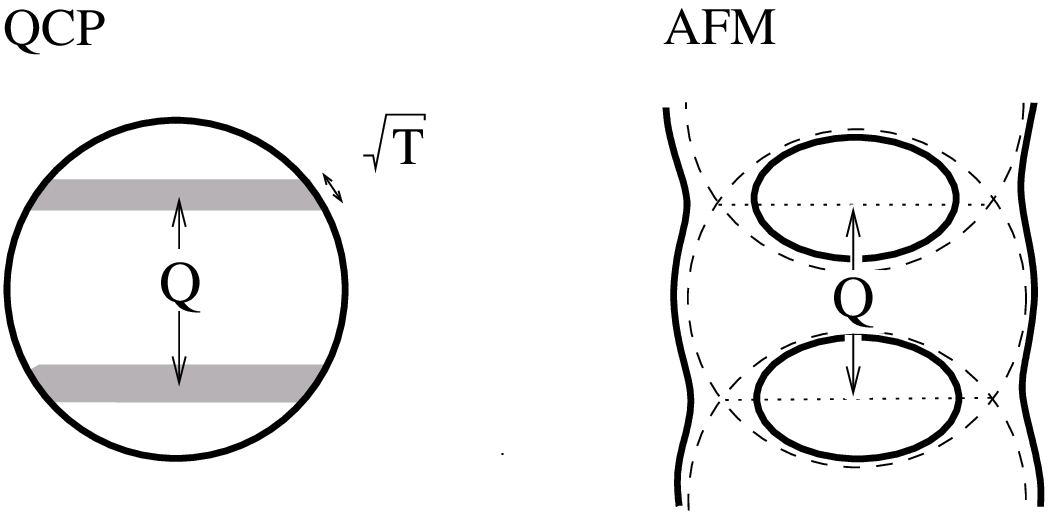}{
Quantum spin density wave
scenario, where the Fermi surface ``folds'' 
along lines separated by
the
magnetic 
$Q$ vector, pinching off into two separate Fermi surface sheets.
}{fig4}
\noindent There are then two major difficulties with the QSDW scenario 
for the heavy fermion QCP: 
\begin{enumerate}

\item {\bf  No breakdown of the Fermi liquid} Away from the hot lines, the Fermi
surface and Landau quasiparticles remain intact at the
QCP.  Thus the specific heat and typical quasiparticle mass do not diverge
but exhibit a weaker singularity, $C_{V}/T= \gamma_{o}
-A\sqrt{T}$ in the QSDW picture\cite{millis}.

\item {\bf  No $E/T$ scaling } The quantum critical behavior predicted by this model has been extensively
studied\cite{hertz,millis}.
In the 
interaction $V_{eff} ({\bf q},\omega)$
the momentum dependence enters
with twice the power of the frequency, so
\[
\tau\sim\xi^z,\qquad (z=2).
\]
In the renormalization group (RG) treatment\cite{hertz} 
time counts as $z$ space dimensions so 
the effective dimensionality is
$D_{eff}=d + z = d+2$. The upper critical dimension is set by
$D_{eff}=4$, or $d_{u}=2$\cite{millis}, so 3D
quantum spin fluctuations will not lead to $E/T$ scaling. 
In three dimensions, QSDW theory predicts that the scale
entering into the energy dependent response functions should scale
as $T^{3/2}$, with a non-universal prefactor\cite{sachdevbook}. 
\end{enumerate}


\subsubsection{Can a 2D spin fluid cure the difficulties?}\label{}

One explanation of $E/T$ scaling and the logarithmically divergent 
specific heat\cite{rosch} is to suppose that the spin fluctuations 
form a \underline{quasi-two-dimensional} 
spin fluid\cite{rosch,mathur}, lying at the critical dimension. 
Inelastic neutron scattering experiments on $CeCu_{6-x}Au_{x}$,
(x=0.1) 
support a kind of reduced
dimensionality in which the 
critical scattering is concentrated
along linear, rather than at point-like regions in reciprocal
space\cite{schroeder,rosch}. More recent data\cite{fak}
may support quasi-2D spin fluctuations at intermediate scales in
$CeGe_{2}Ni_{2}$.

\subsection{New Approaches }

Unfortunately, quasi-two dimensionality still cannot explain
the anomalous exponents in the $E/T$ scaling, and we are essentially 
forced to consider the possibility of 
a fundamentally new {\it interacting} fixed 
point. Two approaches have emerged:

\subsubsection{Local Quantum Criticality}

The momentum-independent scaling term in the inverse dynamic
susceptibility (7)
suggests that the critical behavior associated with the heavy fermion QCP
contains some kind of {\sl local} critical excitation\cite{schroeder}. 
One possibility, is that this 
critical excitation is the spin itself, which would then presumably
develop 
a slow power-law decay\cite{sachdev,sengupta}
\begin{equation}\label{}
\langle S (\tau )S (\tau ')\rangle =\frac{1}{(\tau -\tau ')^{2-\epsilon
}},
\end{equation}
where $\epsilon \ne 0$ signals non- Fermi liquid behavior. 

\begin{figure}[here]
\[ 
\figx{\fight}{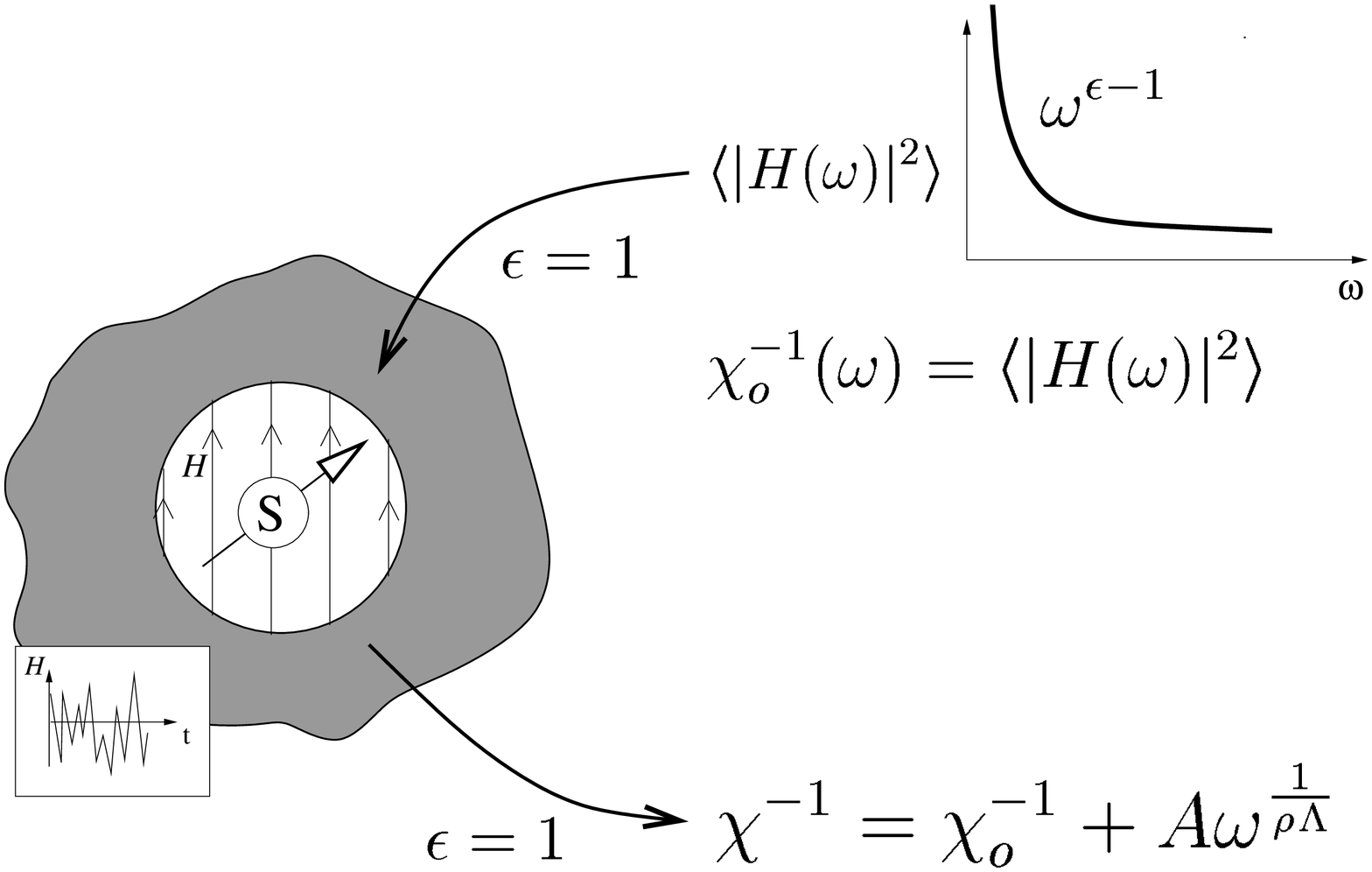}
\] 
\vspace*{-4mm} \caption{\label{fig5} \small In the local quantum
critical theory, each spin behaves as a local moment in a fluctuating
Weiss field. In the theory of Si et al[39], a self-consistent solution
can be obtained for $\epsilon=1$ in which the local susceptibility
develops a self-energy with a non-universal exponent. 
$ M (\omega)\propto \omega ^{\frac{1}{\rho \Lambda}}$.}\end{figure}

Si\cite{si} et al have developed this idea, proposing 
that the
{\it local } spin susceptibility
$\chi_{loc}=\sum_{\vec{q}}\chi (\vec{q},\omega)\vert_{ \omega=0}$
diverges at a heavy fermion QCP, 
From (\ref{lab1}), 
\begin{equation}\label{}
\chi _{loc} (T)\sim \int d^{d }q \frac{1}{({\bf q}-{\bf  Q})^{2} +
T ^{\alpha
}}\sim T^{(d-2)\alpha /2}
\end{equation}
so a divergent local spin susceptibilty requires 
a spin fluid with $d\leq 2$.  Si et al are thus motivated 
to propose that the 
non-trivial physics of the heavy fermion QCP is driven by
the formation of a two-dimensional spin fluid.  
Si et al consider an impurity spin within an effective medium
in which the local Weiss field $H$ has 
a critical power-spectrum (fig. 5.)
\begin{equation}\label{}
\langle \vert H (\omega)\vert ^{2}\rangle \equiv \chi _{0}^{-1}
(\omega)= \omega^{\gamma}
\end{equation}
where $\epsilon$ is self-consistently evaluated using
a dynamical mean-field theory, where  $q-$
dependence of self-energies is dropped. In principle, the method
solves the dynamical spin
susceptibility of the impurity 
$\chi^{-1} 
(\omega)=\chi _{o}^{-1} (\omega)+M (\omega)$. This, in turn
furnishes a ``spin
self-energy'' $M (\omega)$ used to determine 
the spin susceptibility of the medium 
$\chi^{-1} (\vec{q},\omega)= J ( \vec{q})+ M(\omega)$.

Si et al find that a self-consistent solution is obtained for 
$\epsilon=1$,  {\sl if} the spin-self energy 
contains a separate power-law
dependence $M(\omega)\sim \omega^{\alpha }$ 
with an exponent $\alpha = 1/\rho \Gamma$ 
which is determined by the density of states $\rho$
and band-width $\Lambda$ of the bond-strengths in the
two-dimensional spin fluid.
Although self-consistency requires a new power-law in the spin 
susceptibility, 
independent solutions
of the impurity model have not yet shown 
that this feature is indeed  generated by a critical Weiss field. 
This theory nevertheless 
raises many interesting questions:

\begin{enumerate}
\item Is the requirement of a two dimensional spin fluid 
consistent with experiment?  For example- does the
the cubic (and hence manifestly three dimensional) quantum critical
material, $CeIn_{3}$ display a divergent specific heat?

\item If the spin-fluids are quasi-two dimensional, do we expect an
ultimate cross-over to a
three-dimensional QSDW scenario? 

\item If $\alpha $ is non-universal, why are the 
critical exponents in $CeCu_{6-x}Au_{x}$
and $YbRh_{2}Si_{2-x}Ge_{x}$ so similar?

\item 
What stabilizes the local quantum criticality against intersite couplings?

\end{enumerate}

\subsubsection{ Towards the critical Lagrangian.}


If the heavy fermion QCP is a truly three-dimensional phenomenon, 
then a different approach is needed- we need to search for a new class of
critical Lagrangian with $d_{u}>3$\cite{susy}. 
On general grounds, the existence of a Fermi liquid  in the paramagnetic phase
tells us that it must 
find expression in terms of 
the quasiparticle fields $\psi $ in the Fermi liquid, 
\begin{equation}\label{lag}
L = L_{F}[\psi ] + L_{F-M}[\psi,M] + L_{M}[M].
\end{equation}
where $L_{F}$ describes the 
heavy  Fermi liquid,  far from the magnetic
instability, 
$L_{M}$ describes the magnetic excitations that 
emerge above the energy scale $T_{F}^{*} (P)$. 

$L_{F-M}$ describes
the way that the quasiparticles couple to and decay into critical 
magnetic modes; it also 
determines the type of transformation which takes place in the 
Fermi surface which takes
at the QCP. This last point follows because 
away from the QCP, magnetic fluctuations can be ignored in the
ground-state, so that 
$L_{M}\rarrow 0$. In the 
paramagnetic phase, $\langle M \rangle =0$ so $L_{FM}\rightarrow 0$, but in the
antiferromagnetic phase $\langle M \rangle \neq 0$, i.e. 
\[
{{L}}_{\hbox{eff}}= \left\{ 
\begin{array}{cl}
{L} _{F}^{*}[\psi ]&\ \hbox{paramagnet}\cr \cr
{L}_{F}^{*}[\psi ]+{L}_{FM}[\psi , \langle M\rangle ]
&\  \hbox{a.f.m.}\end{array} \right.
\]
where the asterix denotes the finite renormalizations 
derived from zero-point fluctuations in the magnetization.  

If the staggered magnetization
is the fundamental critical field, then we 
are forced
to 
couple the magnetic modes
directly to the spin density 
of the 
Fermi liquid 
\begin{eqnarray}\label{weak}
L_{F-M}^{(1)}&=& g \sum_{{\bf k},{\bf q} } 
\psi \dg_{\vec{k}-\vec{q}} \vec{\sigma}\psi _{\vec{k}}\cdot \vec{M}_{{\bf q}}. 
\end{eqnarray}
But once the staggered magnetization condenses, this leads 
directly back to a static spin density wave (Fig 4). 

An alternative 
possibility is suggested by the observation that the magnetism
develops spinorial character in the heavy fermi liquid. 
The
Luttinger sum rule\cite{luttinger} governing the Fermi surface volume $V_{FS}$
``
counts'' both the electron density $n_{e}$
\underline{and} the number of 
the number of local moments per unit cell
$n_{{spins}}$\cite{martin,oshikawa}
: \begin{equation}\label{}
2\frac{{\cal V}_{FS}}{(2\pi)^{3}}= n_{e} + n_{spins}.
\end{equation}
The appearance of the spin density in the Luttinger sum rule
reflects the composite nature of the heavy quasiparticles, formed
from 
bound-states between local moments and high energy electron
states. 
Suppose the spinorial character of the magnetic  degrees of freedom seen in the
paramagnet {\sl also } manifests itself
in the decay modes of the heavy quasiparticles.  This would imply that
at the QCP, the staggered magnetization factorizes 
into a spinorial degree of freedom
$\vec{M} (x) = z\dg(x)\vec{\sigma }z (x) $, where $z$ is a 
two-component spin $1/2$ Bose field.  ``Spinorial magnetism''
affords a direct coupling between the magnetic spinor $z$
and the heavy electron quasi-particles via an inner product, 
over the spin indices
\begin{equation}\label{poss2}
L_{F-M}^{(2)} = g \sum _{{\bf k}, {\bf q}}[ z\dg _{{\bf k}-{\bf
q}\sigma}
\psi _{{\bf
k}\sigma }
\chi _{\bf q} \dg 
  +\hbox{H.c}], 
\end{equation}
where conservation of exchange statistics obliges us to 
introduce
of a spinless charge $e$ fermion $\chi $. 
This 
would imply that the composite heavy electron  decays
into a neutral ``spinon''and a spinless charge e fermion 
$e^{-}_{\sigma }\rightleftharpoons s_{\sigma } + \chi ^{-}$.
From this perspective, the heavy fermion QCP involves
spin-charge separation, and the critical Lagrangian 
is a gauge theory.

To go beyond this general discussion we need 
to answer various questions:
\begin{itemize}

\item 
Can we connect
the appearance of local criticality with an underlying
gauge symmetry? 

\item What is the link between the heavy fermion QCP and metamagnetism?

\item Is there a fundamental reason why the heavy fermion QCP exhibits
$H/T$ scaling (e.g as opposed to $H^{\delta }/T$ scaling) ?

\item What feature in the critical Lagrangian can push the upper-critical
dimension above 3? 

\end{itemize}

On the experimental front, 
Hall constant measurements may provide a good way
to discern between the spin density wave and composite quasiparticle
alternative. In the former,
regions around the hot-line do not contribute
to the Hall conductivity, and the change in the Hall constant is 
expected to evolve as the square of the staggered
magnetization\cite{questions}.  By contrast, the composite fermion scenario leads
to a much more rapid evolution: depending on whether the density
of spinless fermions is finite at the QCP the 
Hall constant which either jumps or evolves linearly with
the staggered magnetization\cite{questions}. 
\begin{equation}\label{hall}
\Delta R_{H}\propto  \left\{\begin{array}{cr
}
M_{\bf Q}^{2}, & \qquad \hbox{(vectorial )}
\cr
O (1),  M_{\bf Q}&(\hbox{spinorial})
\end{array} \right.
\end{equation}
The only available Hall measurement at a QCP to date shows a 
change in sign takes place in the close vicinity of the QCP
in critical $CeCu_{1-x} Au_{x}$, 
it is not yet clear whether there is a discontinuity 
at the transition\cite{onuki}.
This is clearly an area where more experimental input is highly
desireable. 

\subsection{Summary}

This paper has discussed the origin of the mass divergence
at the heavy fermion quantum critical point, emphasizing that a
quantum  spin density wave  picture can not explain the observed
properties. The proposal of fundamentally new kinds of
quantum critical point have been reviewed. This is clearly an area
with a huge potential for progress both on the experimental, and 
theoretical front.


\end{document}